\documentclass[12pt,a4paper]{article}

\begin{document}

\title{Reflections on three debated issues\\
{\normalsize }}
\author{Ll. Bel\thanks{e-mail:  wtpbedil@lg.ehu.es}}

\maketitle

\begin{abstract}

A time-dependent model of space-time is used to suggest simple explanations for the increasing of the Astronomical unit, as well as the increasing of the distance from the Earth to the Moon. The Pioneer anomaly is also considered.

\end{abstract}

\section*{Introduction}

The first section gives a short presentation of the concept of rigid space-time models that allow as well as for stationary models to have unambiguous definitions of space and time.

The second section discusses a particular time-dependent spherically symmetric model and derives from it the relevant equations to be used in the last section where the program outlined in the abstract is made more explicit.

We use the following convention: $G=c=1,\ \alpha,\beta \ ...=0,1,2,3 \ i,j,...=1,2,3$,


\section{Rigid space-time models}

Let:

\begin{equation}
\label{1.1}
ds^2=g_{\alpha\beta}(x^\rho)dx^\alpha dx^\beta
\end{equation}
be the line-element of a space-time model. And let $\xi^\alpha(x^\rho)$ be a time-like vector field.

\begin{itemize}
\item  $\xi^\rho$ is the generator of an isometry, i.e. a Killing vector field,  if:

\begin{equation}
\label{1.2}
L(\xi^\rho)g_{\alpha\beta}=0
\end{equation}
where $L(\xi^\rho)$ is the Lie derivative operator.

\item  $\xi^\rho$ is the generator of a Born-rigid motion if:
\begin{equation}
\label{1.3}
L(\xi^\rho)\hat g_{\alpha\beta}=0
\end{equation}
where:

\begin{equation}
\label{1.4}
\hat g_{\alpha\beta}=g_{\alpha\beta}+u_\alpha u_\beta,
\quad u_\alpha={\xi}^{-1}\xi_\alpha, \quad \xi=+\sqrt{|g_{\alpha\beta} \xi^\alpha\xi^\beta|}
\end{equation}
Born rigid motions generators are defined up to an arbitrary multiplicative function, thus leaving time associated concept fully ambiguous.

\item We shall say that $\xi^\rho$ is the generator of a rigid motion if:

\begin{equation}
\label{1.5}
L(\xi^\rho)\bar g_{\alpha\beta}=0, \ \ \ L(\xi^\rho)\bar u_\alpha=0,
\end{equation}
where:

\begin{equation}
\label{1.6}
\bar g_{\alpha\beta}=\xi^2 \hat g_{\alpha\beta}, \quad  \bar u_\alpha=\xi^{-1}u_\alpha,
\end{equation}

\end{itemize}
Isometry and rigid motions generators are defined up to a constant factor.

Choosing one factor and using an adapted system of coordinates  such that:

\begin{equation}
\label{1.7}
\xi^0=1, \quad \xi^i=0,
\end{equation}
the line-element (\ref{1.1}) becomes:

\begin{equation}
\label{1.8}
ds^2=-\xi^2(t,x^k)(-dt+\varphi_i(x^k) dx^i)^2+\xi^{-2}d\bar s^2,\  d\bar s^2=\bar g(x^k) dx^i dx^j
\end{equation}
where $\xi$ is the only function that can depend on $t$. The restricted adapted coordinate's transformations are:

\begin{equation}
\label{1.9}
t^\prime=kt+\psi(x^j), \ x^{\prime i}=\Phi^i(x^j)
\end{equation}
and therefore $t$ is defined up to a shift of origin along each trajectory of $\xi$.

As it is obvious, every generator of an isometry is also a generator of a rigid motion.  On the other hand, among the space-time models that posses rigid motion generators that are not isometries we can mention all Robertson-Walker ones, (See Ref. \cite{Bel}) because their line elements can be brought from their usual form to the form (\ref{1.8}) with an appropriate time transformation.

Very important for this paper is the fact that if the space-time rigid model is not stationary then it is not possible to choose the factor of $\xi^\rho$  so that the time  $t$ coincides with the proper time $\tau$ all along a prescribed trajectory of $\xi^\rho$. The better that can be done is to have $t$ and $\tau$ to coincide at a single event of the prescribed trajectory. Thus $t$ is a well defined time but its relationship with $\tau$ is not operationally simple.

It is noteworthy that rigid models of space-times share with stationary ones the fact that if the coordinates $x^i$ are harmonic coordinates of $d\bar s^2$ then they are also harmonic coordinates of the space-time $ds^2$. This follows from:

\begin{equation}
\label{1.10}
\partial_j(\sqrt{|g|}g^{jk})=\partial_j(\sqrt{|\bar g|}\bar g^{jk}) \ \  \hbox{and}\ \
\partial_t(\sqrt{|g|}g^{0k})=0
\end{equation}
The first of these equations follows from a simple algebraic relationship between the coefficients of $ds^2$ and $d\bar s^2$, while the second follows  from the second equation (\ref{1.5}).

Noteworthy is also the fact that when the rigid motion generator is i-rotational, i.e. when

\begin{equation}
\label{1.11}
\partial_i\varphi_j-\partial_j\varphi_i=0
\end{equation}
in which case (\ref{1.9}) allows to transform $\varphi_i$ to zero, then, using a diagonal form of ${d\bar s}^2$:

\begin{equation}
\label{1.11.1}
{d\bar s}^2=F_i(x^j){dx^i}^2
\end{equation}
three of Einstein's tensor components, namely $S^0_i$, become the very simple expressions:

\begin{equation}
\label{1.12}
S^0_i=\partial_{it}\xi^{-1}
\end{equation}
so that in this case and in vacuum we necessarily must have:

\begin{equation}
\label{1.12.1}
\xi^{-1}=f_t(t)+f_x(x^i)
\end{equation}


\section{Spherically symmetric i-rotational approximate models}

Let us consider the following rigid spherically symmetric space-time
model whose line-element is:

\begin{equation}
\label{1.12.2}
ds^2=-A^2dt^2+A^{-2}d\bar s^2, \quad d\bar s^2=M^2 dr^2+N^2r^2d\Omega^2
\end{equation}
where to start with we assume that:

\begin{equation}
\label{1.13}
A^{-1}=\left(1-\frac{2m}{r}\right)^{-1/2}+(1+2pt)^{1/2}-1
\end{equation}

\begin{equation}
\label{1.14}
M=\frac{1}{\sqrt{1+p^2r^2}},\ \ \ N=\sqrt{1-\frac{2m}{r}}
\end{equation}
If $p=0$ then (\ref{1.12.2}) is the Schwarzschild model, $r$ being the curvature coordinate. If $m=0$ then (\ref{1.12.2}) is Milne's flat model of a universe but $t$ is not the global proper time that the model allows to use. From Birkoff's theorem we know that these are the only two vacuum models that could be considered.

Nevertheless, as it stands, (\ref{1.12.2}), without assuming $p=0$ or $m=0$, can be considered a meaningful vacuum approximate solution of Einstein equations as long as $\epsilon_1=m/r$, $\epsilon_2=pr$ and  $\epsilon_3=pt$ are small enough compared to 1 so that their squares and products can be neglected in a relevant domain of space during appropriate time intervals $[0,t]$, $t=0$ corresponding to the epoch when the Newton constant has been measured (See below). In fact with these assumptions the non zero components of the Einstein tensor are:

\begin{equation}
\label{1.15}
S^0_0=4\frac{m}{r^3}\epsilon_2^2
\end{equation}

\begin{equation}
\label{1.16}
S^1_1=-2\frac{m}{r^3}(\epsilon_2^2+\epsilon_1\epsilon_3)
\end{equation}

\begin{equation}
\label{1.17}
S^2_2=S^3_3=-\frac{m}{r^3}(\epsilon_2^2-2\epsilon_1\epsilon_3)
\end{equation}
 The common factor $m/r^3$ in these equations giving the order of magnitude of the Rieman tensor, the $\epsilon$ factors measure the quality of the approximation. It is easy to improve somewhat this approximation but this is unnecessary to our purpose.

With the preceding approximation, that we shall use from now on, often without further mention, and sometimes dropping some of the smaller terms, we obtain the following results:

\begin{itemize}

\item At any given location $r=r_e$ the flow of proper-time $\tau$ with respect to the local coordinated time $t$ is governed by the differential relationship $d\tau=Adt$ so that at the approximation considered we have:

\begin{equation}
\label{1.18}
d\tau=\frac{1}{\sqrt{1+2pt}}\left(1-\frac{m}{r\sqrt{1+2pt}}\right)dt
\end{equation}
that for short time intervals can be approximated to:

\begin{equation}
\label{1.19}
d\tau=(1-pt)dt \quad \hbox{or} \quad \frac{d^2\tau}{dt^2}=-p
\end{equation}
along a particular world-line with radial $r=r_e$ trajectory. This can be done choosing appropriately the factor $k$ in (\ref{1.9}).

\item The force per unit mass on a test particle at rest at a coordinate distance $r$ from the center is :

\begin{equation}
\label{1.17.1}
F(r)\equiv \frac{d^2r}{d\tau^2}=-\Gamma^r_{tt}\left(\frac{dt}{d\tau}\right)^2
\end{equation}
or:
\begin{equation}
\label{1.17.2}
F(r)=-\frac{m}{r^2(1+2pt)^{3/2}}
\end{equation}
This can be interpreted as a time dependence of Newton's constant in which the force per unit mass would be independent of the velocity of the particle derived from a time and position potential::

\begin{equation}
\label{1.17.3}
F_i(x^k)=-\partial_iA^2(t,r(x^k))
\end{equation}
from where it follows the necessity of specifying the time at which any variable universal constant has been measured.
 
\end{itemize}


\section{Secular increasing of the astronomical unit}

Let us consider the Newtonian-like theory described by (\ref{1.17.3}). By this we mean that we consider the equations of motion of a test particle to be in polar coordinates:

\begin{equation}
\label{1.20}
\ddot{r}-r\dot{\varphi}^2=-\frac{m}{r^2}(1-3pt)
\end{equation}
and:

\begin{equation}
\label{1.21}
2\dot{r}\dot{\varphi}+r\ddot{\varphi}=0 \ \ \hbox{or} \ \ \ r^2\dot{\varphi}=L
\end{equation}
$L$ being a constant.
We see from these equations that because of the time dependence of the r-h-s of (\ref{1.20}) $r$ can not be constant since $\ddot{r}=0,\ \dot{r}=0$ and $L\neq 0$ would imply $m=0$ . On the other hand, assuming that $\ddot{r}<<1$ allows to derive an approximation for $\dot{r}<1$. From (\ref{1.20}) with $\ddot{r}=0$ and the second Eq. (\ref{1.21}) we have:

\begin{equation}
\label{1.21.1}
r=\frac{L^2}{m}(1+3pt)
\end{equation}
and therefore:

\begin{equation}
\label{1.22}
\dot{r}=3pr
\end{equation}

If we assume $r$ to be one AU then $p=0.5\times 10^{-20} s^{-1}$ would fit an increasing value of $r$ equal to  7 m per century.

If we assume $r$ to be the distance from the center of the Earth to the moon then $p=10^{-19} s^{-1}$ would fit an increasing value of $r$ equal to 4 mm  per year.

Both: the increasing of the AU and the distance from the center of the Earth to the moon are now subject to debate (See Refs. \cite{Williams}, \cite{Claus}and \cite{Iorio}).


\section{The Pioneer anomaly}

 We consider an observer $O$ who is located at a fixed distance $r_e$ of the center of the spherical source, beyond its surface. We consider also a space-probe $P$ moving freely in a radial direction containing the location of $O$. The observer is monitoring the radar-distance of the probe  and the Doppler shift of the frequency of a radio signal sent to $P$ at proper-time $\tau_e$ when this signal  reaches $O$ at proper-time $\tau_a$ after being reflected by $P$. We shall call  each of these individual processes a round, and we shall call a bunch a continuous set of rounds with origins in a short interval of time between $\tau_e$, and $\tau_e+d\tau_e$. We have to keep in mind as we said before, that because of the allowed time transformations (\ref{1.9}), if $p\neq 0$ it is not possible to choose the $t$ coordinate to coincide with proper-time all along the world-line of the observer $O$. This can be done at a single time only.

An acceleration   of clocks (See Refs. \cite{Anderson} and  \cite{Ranada}) has been suggested as a possible interpretation of the Pioneer anomaly. In the model being described here there is indeed an acceleration of clocks. Namely, the acceleration of proper-time $\tau$ with respect to $t$ given in (\ref{1.19}). But to say that this explains the Pioneer anomaly would require to derive from this remark the formula (15) of \cite{Anderson}. In fact the formula that can be derived (See more details in the Appendix) is:

\begin{equation}
\label{1.24}
\nu_a-v_e(1-2v_{model})=-p\nu_e(\tau_a-\tau_e)
\end{equation}
where $\nu_e$ is the reference frequency of the light emitted at time $t_e$ and $\nu_a$ is the observed frequency of the light received at time $\tau_a$. $v_{model}$ is the frequency that would have been received if $p$ had been zero. The speed of the probe and the gravitational field of the source both contribute to $v_{model}$.

\begin{equation}
\label{1.25}
v_{model}=v\left(1+\frac{2m}{r}\right)
\end{equation}
Choosing $p=H$, the Hubble constant, and using the remarkable approximate formula $a_p=H$ (See Ref. \cite{Sanchez})  brings Eq. (\ref{1.24})) to resemble very much to the formula (15) of \cite{Anderson}. But there is a fundamental difference: the r-h-s of this latter reference is $-\nu_ea_pt$
where $t$ is an interval of some 3.000 days.

Both the subtle acceleration of the proper time that our model predicts and the similarity between  Eq. (\ref{1.24}) and (15) of \cite{Anderson} suggest the possibility of some mismatch between the theoretical concepts that have been used and the operational quantities that have been measured. If this is not the case, and since a model based on a local cosmological constant $\lambda$ is likely to be excluded also (See Ref. \cite{Claus 2}), our feeling is that there is no much hope to explain the anomaly in the framework of General relativity (See \cite{Ranada},\cite{Ranada 2}) to compare with a similar point of view.


\section*{Appendix}

If $p(\tau)$ is short enough then we can use the approximation:

\begin{equation}
\label{A.24}
\tau=\frac12 pt^2
\end{equation}

Let us consider a particular bunch of rounds starting at time $\tau_e$ corresponding to the coordinated time $t_e$. To calculate $t_a$ we have to integrate the following differential equation from $O$ to $P$ and back from $P$ to $O$:

\begin{equation}
\label{A.25}
\frac{dr}{dt}=A^2
\end{equation}
that can be approximated to first order as:

\begin{equation}
\label{A.26}
\frac{dt}{1+2pt}=\left(1+\frac{2m}{r}\right)dr
\end{equation}
We obtain thus successively, $r$ being the coordinate location of the probe at the time coordinate $t$:

\begin{equation}
\label{A.27}
t=t_e+r-r_e+p((r-r_e)^2+2(r-re)t_e)+2m\ln\left(\frac{r}{r_e}\right)
\end{equation}

\begin{equation}
\label{A.28}
t_a=t_e+2(r-r_e)+4p((r-r_e)^2+2(r-re)t_e)+4m\ln\left(\frac{r}{r_e}\right)
\end{equation}
from where using:

\begin{equation}
\label{A.29}
\frac{dr}{dt_e}=v\frac{dt}{dt_e}
\end{equation}
where $v$ is the coordinate speed of the probe at time $t$, we get:

\begin{equation}
\label{A.30}
\frac{dt_a}{dt_e}=1+2v+\frac{4mv}{r}+4p(r-re)
\end{equation}
that, using (\ref{1.19}) and $r-r_e=2(t-t_e)$, becomes:

\begin{equation}
\label{A.31}
\frac{d\tau_a}{d\tau_e}=1+2v+\frac{4mv}{r}+p(\tau_a-\tau_e)
\end{equation}
Setting then:

\begin{equation}
\label{A.32}
\frac{d\tau_a}{d\tau_e}=\frac{\nu_e}{\nu_a}
\end{equation}
yields  Eq. (\ref{1.24}) in the main text.

\end{document}